\documentclass[12pt]{iopart}

\usepackage{iopams}
\usepackage{graphicx}
\begin{document}

\paper[Magnetocaloric properties of Fe$_{2-x}T_x$P ($T$ = Ru and Rh)]{Magnetocaloric properties of Fe$_{2-x}T_x$P ($T$ = Ru and Rh) from electronic structure calculations and magnetisation measurements}

\author{$^1$B~Wiendlocha, $^1$J~Tobola, $^1$S~Kaprzyk, $^2$R~Zach, $^3$E~K~Hlil and $^3$D~Fruchart}

\address{$^1$Faculty of Physics and Applied Computer Science, AGH University of Science and Technology, Al. Mickiewicza 30, 30-059 Cracow, Poland}
 
\address{$^2$Faculty of Physics, Mathematics and Applied Computer Science, Cracow University of Technology, Podchorazych 1, 30-084 Cracow, Poland}

\address{$^3$Institut N\'eel, D\'ept. MCMF, groupe IICE, CNRS, BP 166, 38042 Grenoble Cedex 9, France}

\ead{bartekw@fatcat.ftj.agh.edu.pl}
\begin{abstract}
An analysis of the magnetocaloric properties of the pure and substituted Fe$_{2}$P compounds is made based on KKR-CPA electronic structure calculations and magnetisation $M(H,T)$ measurements. The computed electronic densities of states and magnetic moments are used to calculate both the values of the electronic and magnetic entropies, which fairly agree with the experimental findings. To enlighten the magnetic properties above Curie temperature, the paramagnetic state behaviours are simulated using the disordered local moments (DLM) concept. The KKR-CPA computations show, that in Fe$_{2}$P, the Fe magnetic moment of the (3f) site disappears in the DLM state, while the moment of the (3g) site is only little lowered, comparison made with the low temperature ferromagnetic state.

\end{abstract}



\section{Introduction}

Fe$_2$P is a widely-studied compound which orders ferromagnetically (FM) down to $T_C \sim220$ K, correlating two distinct iron sublattices, thus sharing markedly different magnetic moments. The strongest moment ($\sim2.2~\mu_B$) is found on the Fe(3g) pyramidal site formed by 5 P neighbours, whereas the Fe(3f) moment of the tetrahedral (3f) site formed by 4 P neighbours is about four times smaller ($\sim 0.6~\mu_B$). This ferromagnetic-paramagnetic (FM-PM) transition ($T_C$) is accompanied by a very small volume decrease of $\Delta V/V \sim 0.04\%$, without changing the hexagonal symmetry~\cite{roger70}. Here the FM-PM transition is called $"T_C"$ in many references from literature, even if it is not exactly a conventional type Curie temperature. Significant magnetocaloric (MC) properties have been reported recently on this type of compound~\cite{fruchart05}. At 219~K, for a magnetic field ranging from 0 to 1.3 T, the change of magnetic entropy is found $\Delta S_m = 2.5$ J/(kg.K), a value close to that observed for Gd ($3$ J/(kg.K)) at 295~K, a metal used as a reference.

In this paper we focus on pure Fe$_2$P and doped Fe$_{2-x}T_x$P systems, with $T$ being 4d elements as Ru and Rh. Electronic structure as well as magnetic properties of these compounds are derived from the Korringa-Kohn-Rostoker (KKR) method with the coherent potential approximation (CPA). In the case of the ordered Fe$_2$P compound, the CPA is used to study the paramagnetic state, in the framework of so-called disordered local moments (DLM) concept ~\cite{dlm1}. The aim of this work is to determine the MC effect on one hand and the electronic and magnetic properties on the other. Accordingly, the electronic and magnetic contributions to entropy are calculated, since the entropy jump driven by the application of various external magnetic fields appeared to be relevant parameters, thus characterising the MC material.

\section{Computational details}

Electronic structure calculations were performed using the fully charge- and spin-self-consistent KKR method~\cite{kkr99,stopa}. For pure Fe$_2$P, the full potential calculations were done, while for disordered systems, the KKR-CPA technique was used within the muffin-tin model of crystal potential. Fe$_2$P has a hexagonal structure (space group P-62m, No. 189) with the low-temperature lattice 
parameters~\cite{roger70} $a = 5.866$~\AA, $c = 3.456$~\AA. In the unit cell, iron atoms occupy tetrahedral (3f) and pyramidal (3g) sites, while phosphorus is located on (2c) and (1b) sites. It is noticing that in the Fe$_2$P hexagonal type of structure, both the (3f) and (3g) positions approximate rather well a $hcp$ overall metal sublattice. The atomic positions used here are taken from refinements of the neutron diffraction data~\cite{tobola96}, and the muffin-tin radii of 2.4 a.u. (Fe) and 1.7 a.u. (P) were accounted for. As above-mentioned, the CPA approach was also applied to investigate the magnetically disordered system as a model of paramagnetic state (i.e. the DLM state). Bearing in mind that these computations refer to the ground state properties, without accounting for temperature effects, the DLM computations should simulate reasonably magnetic characteristics above Curie temperature. 

Originally, the KKR-CPA-DLM methodology was applied to study the electronic structure of transition metals in paramagnetic state ~\cite{dlm1,dlm2}.  Similarly to the chemical disorder, the magnetic disorder can be seen analogous to that of an $A_xB_{1-x}$ alloy, where atoms $A$ and $B$ keep opposite magnetic moments. In the DLM model, both $A$ and $B$ atoms occupy the same crystallographic site, thus for $x = 0.5$, the total magnetic moment is zero (if the $A$ and $B$ magnetic moments have the same magnitude). In the case of two nonequivalent iron sublattices in Fe$_2$P, this means that for the 3f site atom $A$ is for Fe$^{\uparrow}$(3f) and atom $B$ is for Fe$^{\downarrow}$(3f) and similarly for the 3g site, Fe$^{\uparrow}$(3g) and  Fe$^{\downarrow}$(3g). The "local" $\mu_A$ and $\mu_B$ magnetic moments are non-zero and they are randomly distributed among the crystal sites, as for a ferromagnetic material above $T_C$. The KKR-CPA calculations of the DLM state are fully self-consistent, i.e. both $A$ and $B$ atoms have independent spin-polarised potentials, and consequently the magnetic moments of both atoms are computed independently. As expected, the final magnetic moments in the DLM state are different from those obtained in the ferromagnetic state, since the electronic structure of a magnetically disordered "alloy" is different. Because of the two non equivalent iron sites occupied in Fe$_2$P by "magnetic" atoms, the CPA had to be applied simultaneously on tetrahedral and pyramidal sites. For fully converged crystal potentials, total and site-decomposed spin-dependent densities of states (DOS) were computed, using tetrahedron technique of integration in the reciprocal space~\cite{kaprzyk86}. The total and local magnetic moments were also calculated. The Fermi level ($E_F$) was determined using the generalised Lloyd formula~\cite{kaprzyk90}. 

In order to analyse the MC effect characteristics in the vicinity of the magnetic phase transition, the entropy and its variation was 
considered for. The total entropy can be decomposed into the electronic, magnetic and lattice contributions: 

\begin{equation}\label{eq.ent}
S_{tot} = S_{el} + S_m + S_{lat}.
\end{equation}
The electronic entropy as a function of temperature $T$ can be computed using DOS functions $n(E)$ and the formula~\cite{grimvall}:
\begin{equation}\label{eq.sel}
S_{el} = - R \int_{E_{min}}^{\mu_{c}} dE \ n(E) [f \ln f + (1-f) \ln (1-f)],
\end{equation}
where $R$ is the gas constant, and $f$ is the Fermi-Dirac distribution function, $k_B$ being the Boltzmann constant:
\begin{equation}
f = f(E, \mu_c, T) = \frac{1}{\exp[(E - \mu_c)/k_BT]+1}.
\end{equation}
The chemical potential ($\mu_c$) is a function of temperature $T$, and is obtained self-consistently from DOS normalisation integral, $N_{val}$  
being the total number of valence electrons:
\begin{equation}\label{eq.n}
N_{val} = \int_{E_{bottom}}^{\mu_c} dE \ n(E) f(E, \mu_c, T),
\end{equation}
where integration starts at the bottom of the valence bands. The resulting entropy is given in J/(K mol) per formula unit, when $n(E)$ is normalised to the number of valence electrons per f.u. (note, that the unit cell of Fe$_2$P contains three formula units). 

The electronic entropy weakly depends on external magnetic field, since, in a first approximation, application of an external magnetic field only slightly shifts electronic spectrum. Noteworthy, Fe$_2$P exhibits magnetoelastic first-order phase transition at $"T_C"$, and both the elastic and magnetic transitions are shifted towards higher temperatures under external magnetic field. To some extend, this effect allows studying the impact of a magnetic field on electronic structure near $T_C$ by modifying the crystal structure. Thus, we can assume, that for temperatures slightly above $T_C$ the influence of a magnetic field on the electronic DOS is related to the change of the crystal parameters (from PM to FM ones).

The temperature variation of the magnetic entropy $S_m$ (connected with the magnetic moments degrees of freedom) cannot be easily evaluated from first principles methods, as it would require calculations of magnetic moments variations with temperature.
The DLM approach appears of interest here, since it may provide information on the variation of the magnetic moments due to magnetic disordering (similarly to disordered magnetic structure above $T_C$). However, the DLM system is completely disordered, which is the case for temperatures far above $T_C$, where no short-range ordering takes place. Thus the DLM model appears well appropriated to calculate the magnetic entropy, at relatively high temperatures, only.

In order to investigate the magnetic entropy behaviours close to the transition point one can use the magnetic moments computed in the ferromagnetic state, with the temperature effects taken into account within the mean-field approximation. The magnetic entropy of a ferromagnetic system at a temperature $T$ and in a magnetic field $H$, is given by~\cite{vonsovsky}:

\begin{equation}
S_m = R \left[ \ln\frac{\sinh \frac{2J+1}{2J}x}{\sinh \frac{1}{2J}x} - x B_J(x) \right],
\end{equation}
where $B_J(x)$ is the Brillouin function, and 
\begin{equation}
x = \frac{g\mu_BJH}{k_bT} + 3\frac{T_C}{T}\frac{\mu(T)}{g\mu_B(J+1)},
\end{equation}
where $\mu(T)$ is the effective magnetic moment of atom in temperature $T$, obtained from self-consistent equation 
$\mu(T) =  \mu(0)B_J(x)$. The $\mu(0) = gJ$ is the zero-temperature saturation magnetic moment (in $\mu_B$ units), as obtained from the KKR-CPA calculations. This formula gives the well-known maximum saturation entropy $S_m^{max} = R\ln(2J+1)$. 

For the considered multi-component compounds, the magnetic entropy was calculated independently for each magnetic atom, on 3f and 3g sites respectively, and the atomic contributions were added, since the entropy is an intensive parameter. 
In the present work, we intend to discuss the entropy behaviours, as far as possible, without using any adjustable parameters, like those needed for the Bean-Rodbell model, as e.g. to reproduce the MC properties of MnAs ~\cite{oliveira2}. 
Finally, the experimental ordering temperature $T_C$ is the only external parameter used in the model. 
We are aware that the simplified mean-field approach cannot fully represent complexity of magnetic properties of systems such as Fe$_2$P, but one should find it valuable when discussing trends in magnetic entropy. It is worth noting, that in Fe$_2$P the change of the unit cell volume near the magnetic transition is very small from $\Delta a/a = - 0.06\%$ and $\Delta c/c = 0.08\%$~\cite{roger70}. Thus, as far as the magnetic entropy is considered, the fact, that the transition is of 1$^{\rm st}$ order type, should not significantly affect the obtained results. The effect of the magneto-elastic transition on the electronic structure was checked being negligible, since the computed magnetic moments and DOS’s remained practically unchanged in KKR calculations, based on the crystal structure data measured below and above $T_C$.

The third entropy component, related to the crystal lattice ($S_{lat}$), usually gives the largest contribution to the absolute value of the total entropy.
In 2$^{\rm nd}$ order transition systems it is commonly assumed that $S_{lat}$ does not depend on the magnetic field. 
Thus, for the magnetocaloric analysis, the lattice entropy is not so important, since only the change in the total entropy, caused by the magnetic field, is important, and the magnetic field affects only the magnetic entropy term. However, this assumption generally fails in compounds exhibiting a 1$^{\rm st}$ order transition, because the field induced PM-FM transition is accompanied by an elastic type transition. This behaviour may modify dynamic properties of the system including the lattice entropy. However, as was earlier mentioned, for Fe$_2$P at the PM-FM transition, the unit cell volume changes only slightly, so we may treat the lattice as a background for the magnetic transition.
It was deduced ~\cite{fe2p-transition,fe2p-heat}, that the lattice entropy change along the transition is $\Delta S_{lat}\sim-0.06$ J/(kg.K), to compare with the total change $\Delta S\sim 1.4$ J/(kg.K) determined by specific heat measurements in zero magnetic field. For finite magnetic fields, one can assume that the lattice contribution is not considerably larger.


\section{Experimental details}

Structural and magnetic investigations were performed for samples with composition Fe$_{1.85}$Ru$_{0.15}$P and Fe$_{1.75}$Rh$_{0.25}$P. The samples were prepared starting from 99.9\% pure elements, mixtures of fine powders were sealed in evacuated silica tubes and then progressively heated up to 850$^o$C for 7 days. 
A final overheat treatment was realised by using the HF melting technique in a cold crucible in order to melt and then to anneal the samples within the tubes. 
The quality and crystal structure of the samples were checked by X-ray powder diffraction using a conventional Bragg-Brentano type 
diffractometer equipped with a backscattering graphite monochromator working at $\lambda_{Co} = 1.7902$~\AA. The magnetic measurements were performed 
using the BS1 and BS2 extraction type magnetometers developed at the Institut N\'eel, for high and low temperatures, respectively. 
From the isotherm magnetisation measurements with steps $\Delta T$ as small as 2~K, the magnetic field 
varying from 0 to 3~T, the entropy variation was calculated, using the Maxwell-Weiss relation:
\begin{equation} 
\Delta S(H_\mathsf{max}) = \int_0^{H_\mathsf{max}} \left(\frac{\partial M}{\partial T}\right)_H dH.
\end{equation}

\section{Results and discussion}

\subsection{Fe$_2$P}

\begin{figure}[htb]
\includegraphics[width=0.60\textwidth]{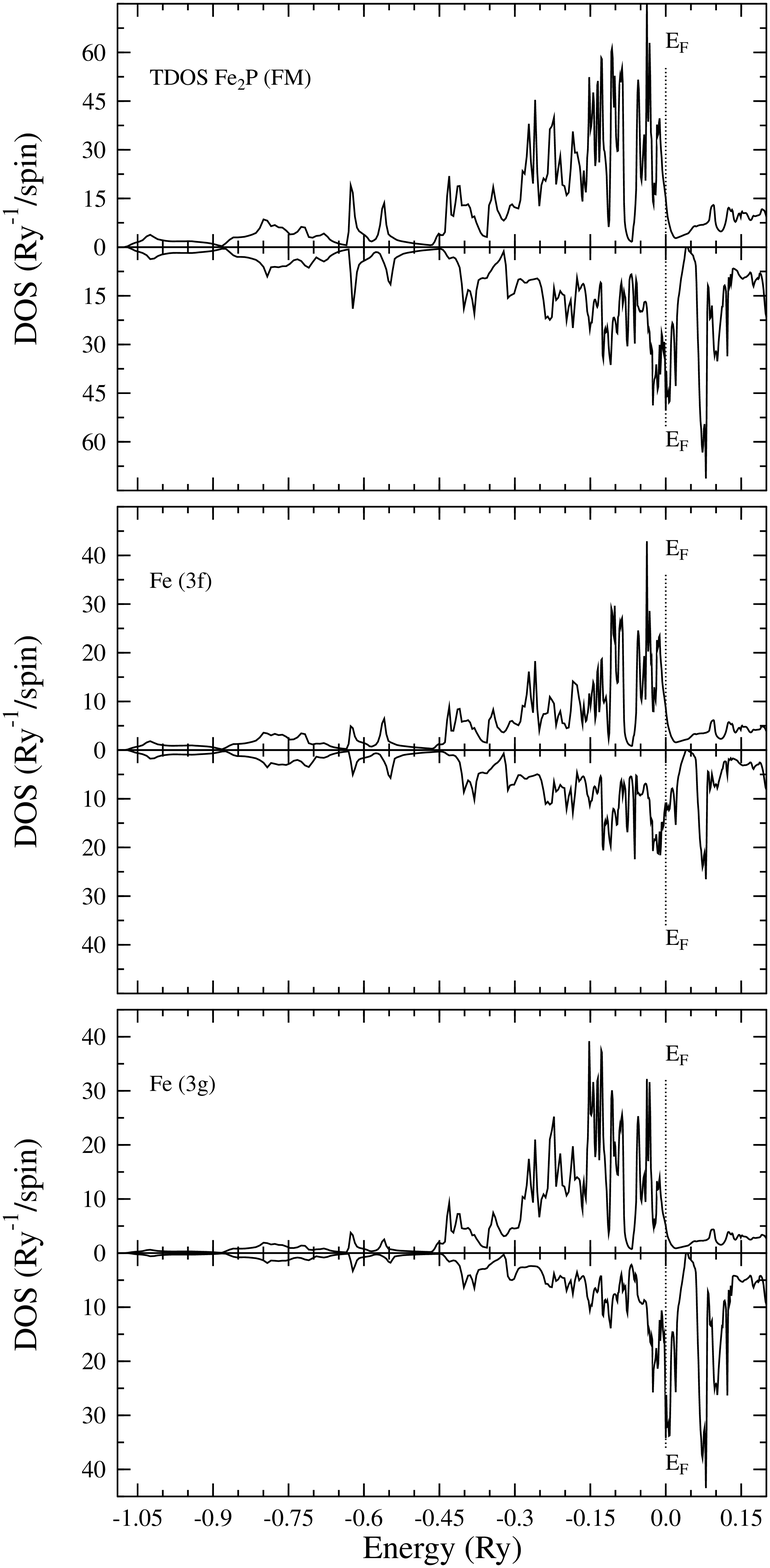}\caption{\label{fig-fe2p-fm}Density of states for Fe$_2$P in ferromagnetic state: total density of states (upper panel), site-decomposed densities for Fe(3f) (middle panel) and Fe(3g) (lower panel).}
\end{figure}

Density of states for ferromagnetic Fe$_2$P is presented in \fref{fig-fe2p-fm}. The lowest lying states are formed from $s-$orbitals of P and $s-$ and $p-$orbitals of Fe, and more important hybridisation of d-states of Fe with p-states of P builds the conduction bands near the Fermi level.  
$E_F$ is located on the decreasing slope of DOS curve for spin-up electrons, while it is confined in the sharp DOS peak for spin-down 
electrons, attributed especially to the Fe(3g) site. 
The high polarisation of Fe(3g) $d-$DOS results in the large value of magnetic moment 2.33~$\mu_B$, comparing to only 0.50~$\mu_B$ computed for Fe(3f). The P atoms possess small negative magnetic moments -0.12~$\mu_B$ (2c) and -0.05~$\mu_B$ (1b). The total magnetisation per formula unit was found to be as large as  $M_{tot} = 2.71~\mu_B$. All these magnetic moments values (obtained from full-potential KKR calculations), well agree with the experimental neutron data ($2.20~\mu_B$ for Fe(3g), $0.60~\mu_B$ for Fe(3f)), as well as the previous magnetisation measurements $M_{tot} = 2.87~\mu_B$ per f.u.~\cite{tobola96}. The magnetic moments resulted from the muffin-tin KKR computations are only slightly different with $2.31~\mu_B$ for Fe(3g), $0.80~\mu_B$ for Fe(3f), and $M_{tot} \sim 3.0~\mu_B$.

\begin{figure}[htb]
\includegraphics[width=0.60\textwidth]{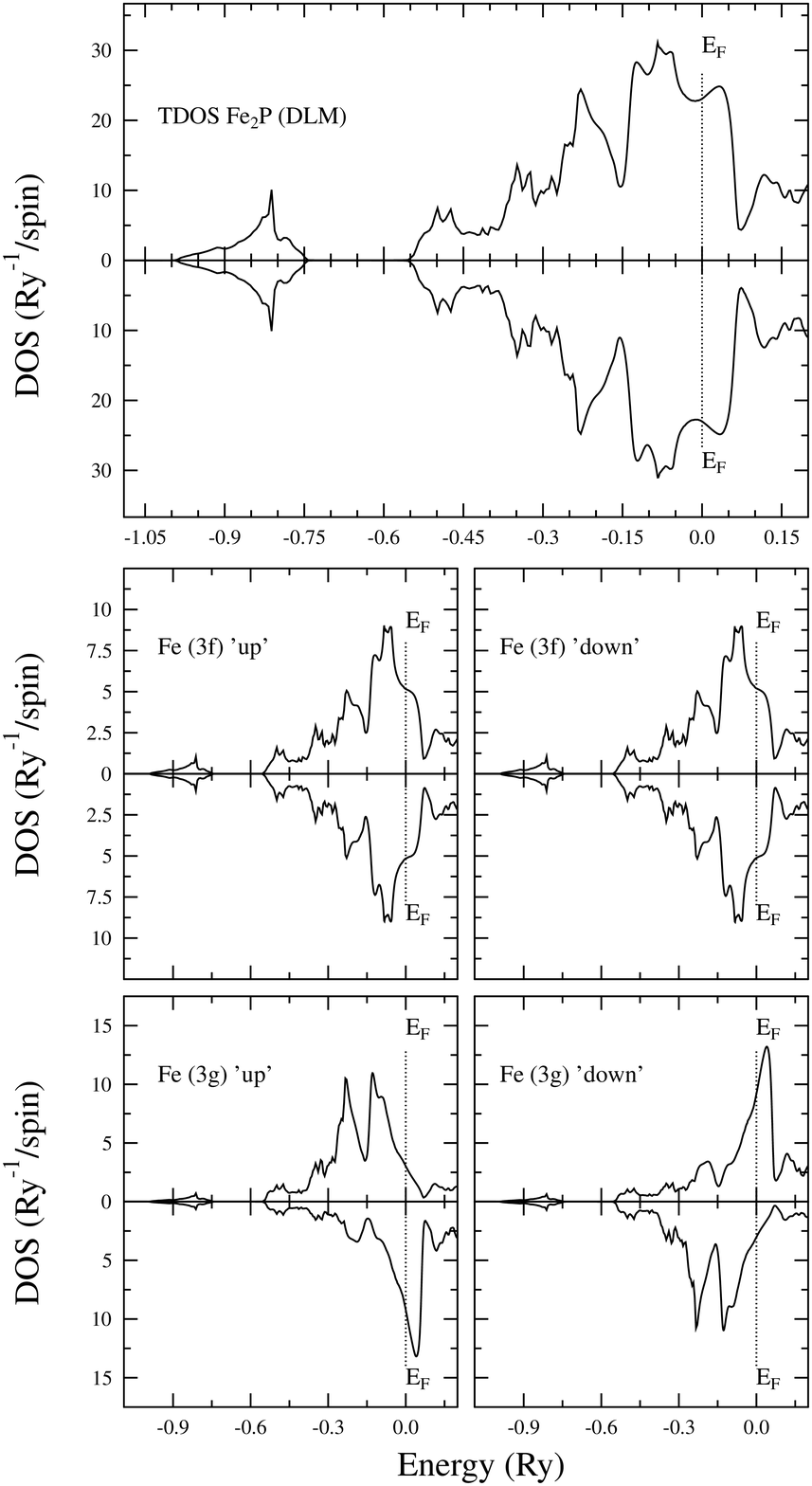}\caption{\label{fig-fe2p-pm}Density of states for Fe$_2$P in disordered local moments state: total density of states (upper panel), site-decomposed densities for Fe(3f) (middle panel) and Fe(3g) (lower panel).}
\end{figure}

\Fref{fig-fe2p-pm} presents the DOS in the "paramagnetic" DLM state. On first glance we can notice a smoothing of the DOS curve with respect to the ferromagnetic state, likely resulted from the magnetic disorder. Identical DOS shapes for both spin directions gives zero magnetisation per unit cell, as expected in the DLM state. 
The most interesting feature of the KKR-CPA-DLM results is related to the vanishing of the local magnetic moment on the Fe(3f) site, since the site-decomposed DOS components 'up' and 'down' do not show spin-polarisation (\fref{fig-fe2p-pm}). 
This suggests, that for the Fe(3f) sublattice the disappearance of local magnetic moments in paramagnetic state may be dramatically fast, as the absence of long-range order appears to be sufficient to destroy magnetic polarisation on this sublattice (even without accounting for temperature effects). 
On the other hand, the Fe(3g) sublattice still exhibits a strong spin-polarisation, and the 'local' magnetic moments are only slightly lowered, comparing to the FM case. 
The spin-opposite magnetic moments are $+2.1$~$\mu_B$ and $-2.1$~$\mu_B$, respectively, giving a total magnetic moment equal to zero per Fe(3g) site. 
Furthermore, a separation of the total DOS into two energy ranges can be noted, a fact which was not observed in the ferromagnetic state. 
The lower part of DOS, similarly to the FM state, is formed from $s-$orbitals of phosphorous and $s-$, $p-$orbitals of iron, but with a lower contribution of the iron states, since these states are transferred towards higher energies. 
Surprisingly, the DOS at $E_F$ in the DLM state is much smaller, than that it is found in the FM one, i.e. $n(E_F) \simeq$ 45~Ry$^{-1}$/f.u (DLM) and 60~Ry$^{-1}$/f.u. (FM). 
The marked decrease of $n(E_F)$ is related with the decrease of DOS on Fe(3g) site, being about 44~Ry$^{-1}$ for the FM phase (mainly spin down) versus only 24~Ry$^{-1}$ for the DLM phase. 
Consequently, the DOS decrease determined near the Fermi level leads to an unexpected decrease of the electronic entropy in the disordered phase. 

It is worth noting, that the situation found here is similar to what we observe for the MnFeP$_{1-x}$As$_{x}$ series of magnetic material, which have been demonstrated sharing among the most important MC properties \cite{bruck02}, furthermore which structure is directly isotype to the Fe$_2$P one~\cite{mnfeasp-bacmann,mnfeasp-malaman}. The magnetic moments in the DLM state on pyramidal site (occupied by Mn) remain large (about 3.0~$\mu_B$) while those on tetrahedral site (occupied by Fe) are lowered twice (from about $1.0~\mu_B$ to 0.5~$\mu_B$).

\begin{figure}[htb]
\includegraphics[width=1.0\textwidth]{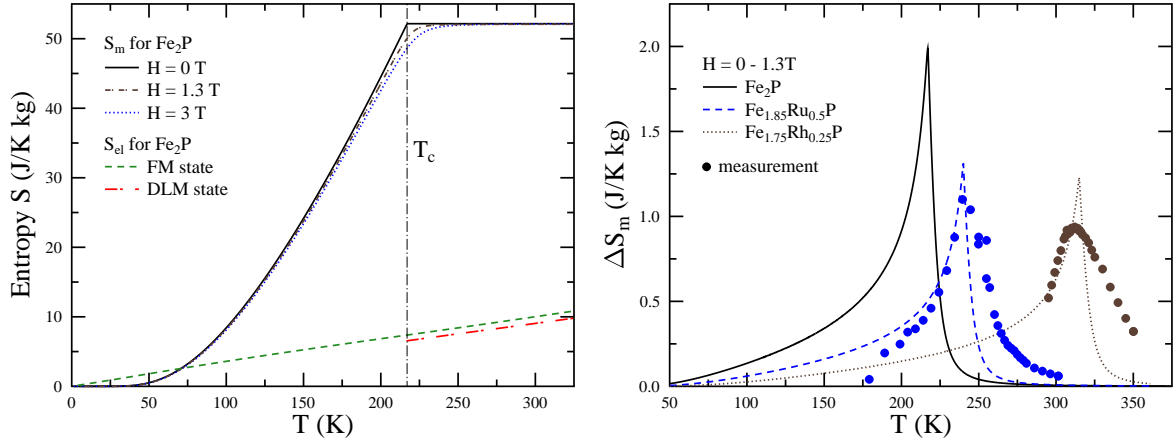}\caption{\label{fig-entr} Magnetic and electronic entropy for Fe$_2$P (left) and magnetic entropy change, induced by field variation between 0 and 1.3~T (right) for Fe$_{2-x}T_x$P.}
\end{figure}

The electronic entropy variation with temperature, calculated using equation \eref{eq.sel} (with chemical potential obtained from equation \eref{eq.n}) and applying the FM and DLM DOS, are shown in \fref{fig-entr}, where DLM is plotted only for $T>T_C$. We can observe a linear increase of the electronic entropy with increasing temperature, leading to a quite noticeable value comparing to the magnetic entropy at low temperatures. 
For $T < 70$~K electronic entropy is even larger, than the magnetic one. For $T = T_C$ the electronic entropy contributions are $S^{FM} = 7.36$~\makebox{\makebox{J/(K kg)}} and $S^{DLM} = 6.55$~\makebox{J/(K kg)}, which would give the change in FM-PM transition $\Delta S_{el} = S^{DLM} - S^{FM} = -0.81$~\makebox{J/(K kg)}, if we assumed that the DOS below and above the magnetic transition correspond to the FM and DLM states, respectively. 
However, this value of $S_{el}$ should be regarded as an upper limit for the electronic entropy change during the transition, since the (DLM) DOS corresponds to temperatures far above $T_C$, and the (FM) DOS is valid at $T = 0$~K. 
Experimentally, the evolution of DOS with temperature is expected to be much smoother, thus the electronic entropy change is expected to be smaller. Besides, the small decrease of the unit cell volume at the first-order transition was found to have a negligible impact on the electronic entropy.

The magnetic entropy as a function of temperature and magnetic field, calculated using KKR values of FM moments, is reported in \fref{fig-entr}. Values of $J$ = 1 and ${1\over2}$ for the (3g) and (3f) magnetic moments, respectively, were assumed, and then $g$ was computed using the equation $\mu(0) = gJ$. One can see that the field-dependent entropy curves, plotted for magnetic field values $H$ of 0~T, 1.3~T and 3~T, increase monotonically and get saturation values above $T_C$~\footnote{The differences, between $S_m$ obtained using moments from the full potential and spherical potential KKR calculations are small, not exceed 5\% for the maximum $S_m$ value, and 0.1~\makebox{J/(K kg)} for the magnetic entropy change $\Delta S_m$ at $T_C$.}. 

The entropy change, defined as $\Delta S_m = S_m(H = 0) - S_m(H > 0)$ is also presented, for a field variation $\Delta H = 1.3$~T. The 
computed value of $\Delta S_m = 2.0$~\makebox{J/(K kg)} fairly corresponds to the measured one $\Delta S_m = 2.5$~\makebox{J/(K kg)}~\cite{fruchart05}. For a larger magnetic field variation $\Delta H = 3$~T, the calculated change of magnetic entropy is about  $3.5$~\makebox{J/(K kg)}. Interestingly, the results for the magnetic entropy and its jump remain almost unchanged (differences $\sim$ 2\%), if instead of fixing $J$, the 
value of $g = 2$ is assumed and then $J$ is calculated. For temperature much above $T_C$, the magnetic entropy contribution decreases because of the decreasing (vanishing) of magnetic moments due to the magnetic disorder, as predicted by our DLM calculations, and also due to temperature effects, both decreasing the amplitude of magnetic moments. The upper limit of high-temperature $S_m$ is about 40~\makebox{J/(K kg)}, as can be estimated using the local DLM magnetic moment of Fe(3g) site ($2.1~\mu_B$).

\subsection{Substituted Fe$_{2-x}T_x$P}

The Fe$_{2-x}Tx$P compounds, with $T$ being the 4$d$-elements Ru and Rh, were also considered owing to the observed and unexpected increase of Curie temperature from T$_C \sim$ 220~K for Fe$_2$P to 240~K for Fe$_{1.85}$Ru$_{0.15}$P and up to 315~K  for Fe$_{1.75}$Rh$_{0.25}$P~\cite{fruchart05,balli-conference}. The crystal structures of those compounds, prepared as mentioned above, were determined using X-ray diffraction. In the case of Fe$_{1.85}$Ru$_{0.15}$P a change of the crystal symmetry from 
hexagonal to orthorhombic (the Co$_2$P-type unit cell) was observed upon Ru substitution \cite{Artigas}. At room temperature the lattice parameters were refined as $a = 5.756$~\AA, $b = 3.579$~\AA, $c = 6.622$~\AA. In the Co$_2$P-type structure, all the atoms are placed in 4c $(x, {1\over4}, z)$ sites and the following atomic positions were determined: Fe-tetrahedral $(0.1571, {1\over4}, 0.4285)$, Fe-pyramidal $(0.01896, {1\over4}, 0.82584)$ and P $(0.2569, {1\over4}, 0.1113)$. The Fe$_{2-x}$Ru$_x$P system maintains the hexagonal unit cell at the lowest Ru concentrations, as previously reported~\cite{ferup-fujii}. 
On the other hand, the Fe$_{1.75}$Rh$_{0.25}$P crystallises in the Fe$_2$P-type unit cell, with the lattice parameters: $a = 5.886$~\AA, $c = 3.486$~\AA, and atomic positions Fe(3f) $(0.26406, 0, 0)$, Fe(3g) $(0.58507, 0, {1\over2})$, P $(0, 0, {1\over2})$, P (${1\over3}, {2\over3}, 0)$. 

The impact of both the relative electronegativity and the atomic radii of the metals $T$ and $T'$ in the $TT'$P series was determined earlier in terms of "selection rules" based on metal-metal interactions~\cite{Robert83}. The largest (smallest) $T$ or $T'$ atom would prefer to occupy the largest (smallest) site, the pyramidal (tetrahedral) one. Conversely, the most (less) electropositive $T$ or $T'$ atom would prefer being coordinated by 5 (4) P neighbours, i.e. pyramid versus tetrahedron. Also, if metal-metal interactions decreases, the polytype structures succeed from tetrahedral (T2: SG P4/$nmn$, Fe$_2$As type) to hexagonal (H3: SG P-62$m$, Fe$_2$P type) and to orthorhombic (O4: SG P$nma$, Co$_2$P type) with a decrease of the unit volume per formula unit ($V_{\rm T2} > V_{\rm H3} > V_{\rm O4}$)~\cite{Artigas,Robert83}. For intermediate conflicting situation between the relative electronegativity and the metal radii, more or less disordered occupation schemes of both the pyramidal and tetrahedral sites have been observed.

\begin{figure}[htb]
\includegraphics[width=1.0\textwidth]{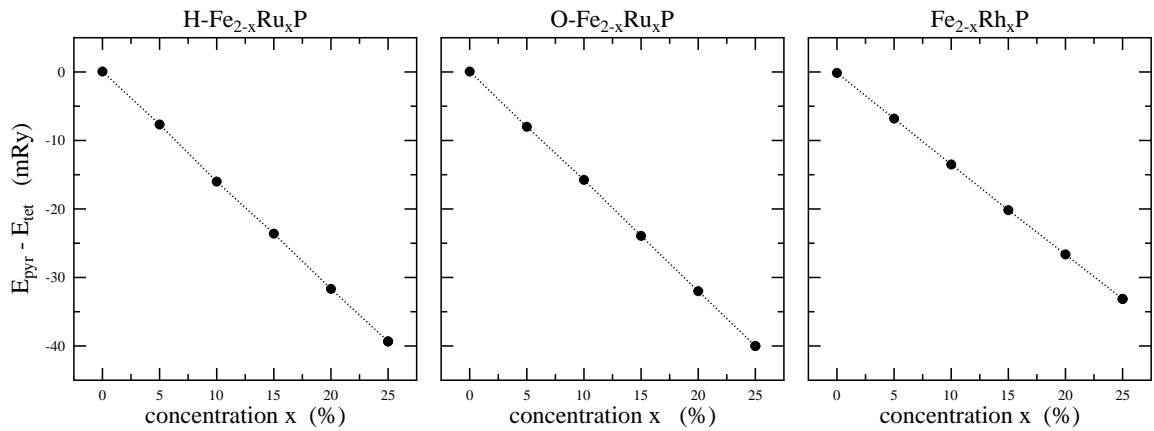}\caption{\label{fig-energy} Total energy difference per formula unit in ferromagnetic state for Fe$_{2-x}T_x$P alloys, for cases where $T$ is diluted either on pyramidal or tetrahedral site. H- and O-Fe$_{2-x}$Ru$_x$P refer to the hexagonal (H) and orthorhombic (O) structures, respectively. The pyramidal site preference is observed in all cases.}
\end{figure}

Since using X-ray diffraction we are not aware (due to the small substitution rate) either tetrahedral or pyramidal site is preferred when substituting Fe by Rh or Ru, total energy KKR-CPA calculations for both possibilities were undertaken to study the selective substitution. \Fref{fig-energy} presents the difference in total energies for the cases where the substitution element was entered exclusively on tetrahedral or pyramidal sites. The preference of the pyramidal site is clearly observed, being energetically more favourable. Moreover, and in agreement with the "selection rules" recalled here above, the symmetry lowering observed with Ru can be interpreted as trends for more effective site ordering between Fe and Ru.  Interestingly, similar KKR-CPA calculations performed for the Fe$_{2-x}$Ni$_x$P system resulted in the opposite site preference, favouring the tetrahedral (3f) position. This results confirms well the earlier neutron diffraction experiments~\cite{fenip-sredniawa} and also agrees with the so-called “selection rules” since Ni is less electropositive element than Fe, and r$_{Fe} > r_{Ni}$. The previous agreement between experimental and theoretical results gives us a confidence to the site-preference predictions in Fe$_{2-x}$Ru$_x$P and Fe$_{2-x}$Rh$_x$P. Consequently, all results presented below were obtained assuming the pyramidal selective occupation by Ru and Rh. 

\begin{figure}[htb]
\includegraphics[width=1.0\textwidth]{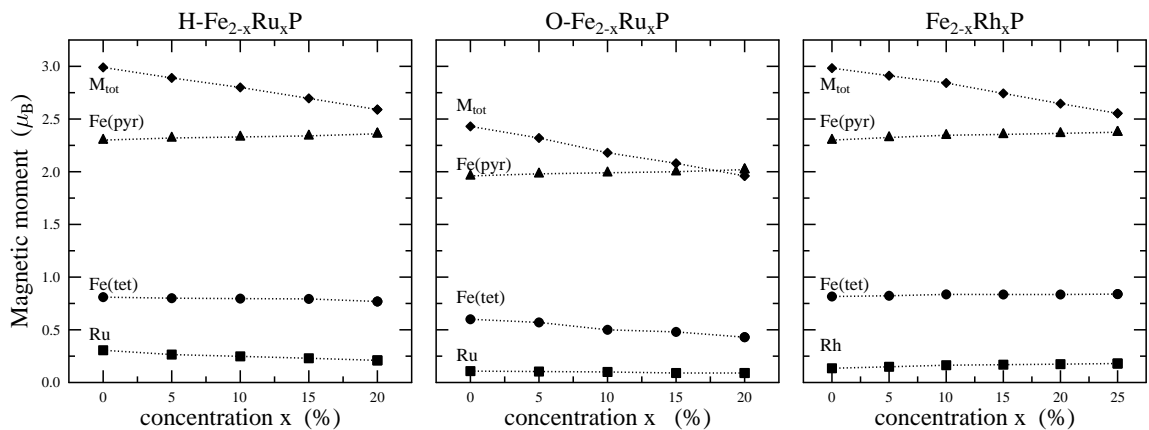}\caption{\label{fig-mom} Magnetic moments per atom and total magnetisation per formula unit in Fe$_{2-x}T_x$P alloys, H- and O-Fe$_{2-x}$Ru$_x$P refer to the hexagonal (H) and orthorhombic (O) structures, respectively.}
\end{figure}

\begin{figure}[htb]
\includegraphics[width=1.0\textwidth]{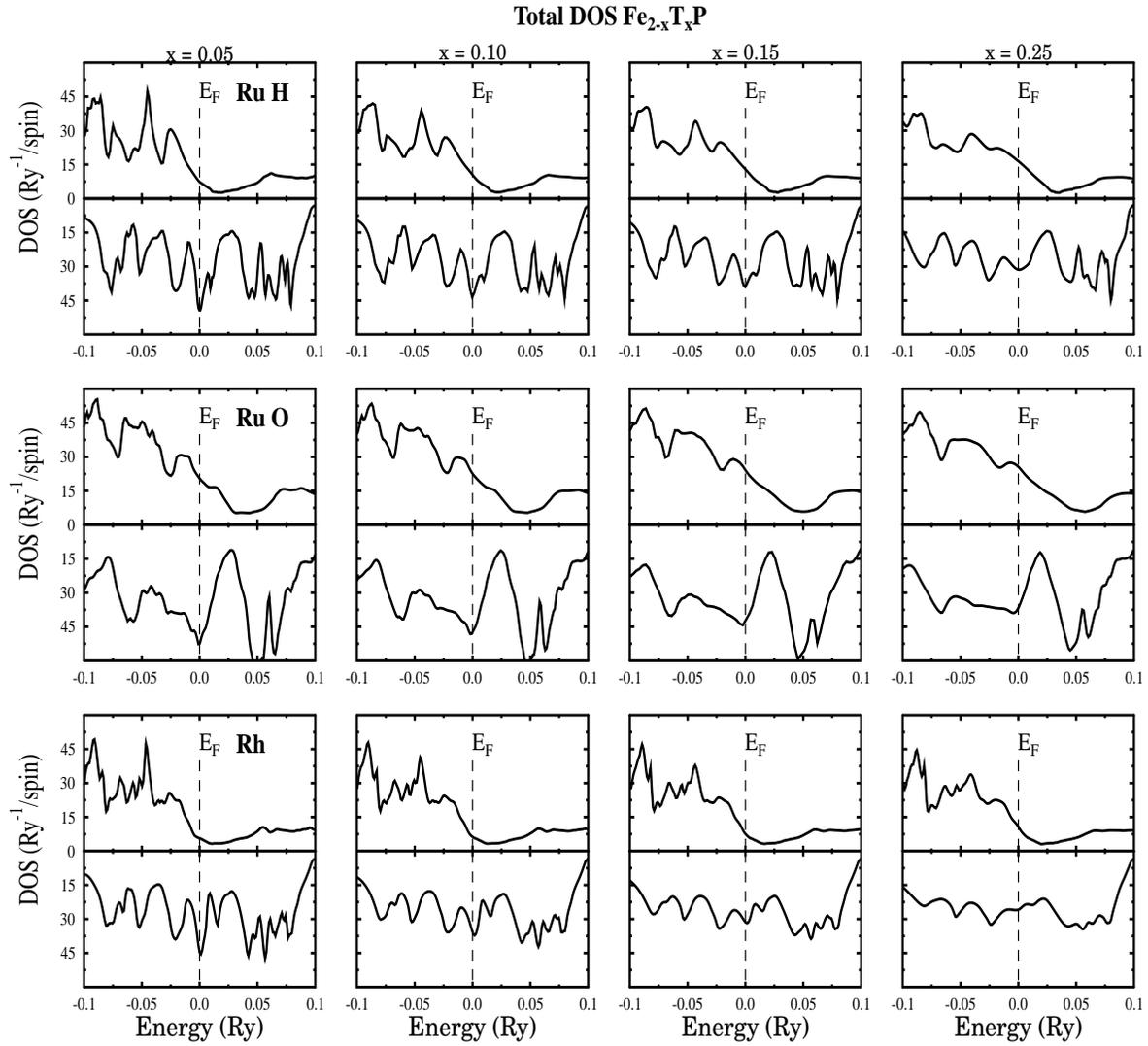}\caption{\label{fig-dos-doped} Evolution of total DOS for Fe$_{2-x}T_x$P alloys. Upper panel: $T$ = Ru assuming the hexagonal (H) structure, middle panel: $T$ = Ru assuming the orthorhombic (O) structure, lower panel: $T$ = Rh.}
\end{figure}

\Fref{fig-mom} shows the evolution of magnetic moments in Fe$_{2-x}T_x$P with the $T$ concentration and \fref{fig-dos-doped} shows the evolution of the corresponding DOS’s. Additionally, to have a better insight in the magnetic behaviours of these compounds, the Fe$_{2-x}$Ru$_x$P system both in hexagonal (correct at low $x$, denoted as 'Ru H') and orthorhombic (denoted as 'Ru O') phases are presented.
In the case of Fe$_{2-x}$Ru$_x$P (Ru is isoelectronic to Fe), the transition to the orthorhombic structure strongly affects the DOS shape, being much smoother, comparing to the hexagonal one. The large DOS polarisation near the Fermi level, which is the characteristic feature of pure Fe$_2$P, practically disappears with Ru substitution. $n$($E_F$) increases (decreases) for spin-up (spin-down) electrons, which leads to the decrease of the total magnetisation, as shown in \fref{fig-mom}. The values of magnetic moment of the constituent atoms change slightly in the considered range of content, but first of all, the orthorhombic phase exhibits much lower magnetisation, comparing to the hexagonal one. For $x = 0.15$ magnetic moments are about: $\mu_{Fe(\rm pyr)} = 2.0~\mu_B$, $\mu_{Fe(\rm tet)} = 0.48~\mu_B$ and $\mu_{Ru} = 0.1~\mu_B$, and are lower with respect to the values expected if the system would preserve the hexagonal structure upon doping (according to a decrease of the unit cell volume $V_{\rm O4} < V_{\rm H3}$).

In the case of Fe$_{2-x}$Rh$_{x}$P the smoothing of spin-up DOS with substitution is less noticeable as seen for Fe$_{2-x}$Ru$_{x}$P, since for $x = 0.15$ we can still observe well pronounced peaks. However, the DOS polarisation and then total magnetisation also decrease, in spite of the fact that the spin-down DOS is less modified. Adding one electron more to the system when replacing Fe by Rh does not significantly move the Fermi level towards the DOS valley (higher energy range), which is in contrast to the rigid-band expectations (see, \fref{fig-dos-doped}). The total magnetisation of Fe$_{2-x}$Rh$_x$P decreases to M$_{tot}$ = $2.5~\mu_B$ per formula unit for $x = 0.25$, due to the atomic moments: $\mu_{Fe(\rm pyr)}$ = $2.37 \mu_B$, $\mu_{Fe(\rm tet)}$ = $0.84~\mu_B$ and $\mu_{Rh}$ = $0.18~\mu_B$. It is worth noting that in Fe$_{2-x}$Rh$_x$P the local moments on both the Fe sublattices increase slightly upon substitution.

The electronic entropy for $T = T_C$ are: $S_{el} = 7.36$ J/(kg.K) and $S_{el} = 10.64$ J/(kg.K) for Fe$_{1.75}$Rh$_{0.25}$P and Fe$_{1.85}$Ru$_{0.15}$P, respectively. Such a difference presumably results from the smaller DOS at $E_F$ for the Rh case 
(see, \fref{fig-dos-doped}). The decrease of magnetisation of these compounds is reflected in their magnetocaloric properties. 

\begin{figure}[htb]
\includegraphics[width=1.0\textwidth]{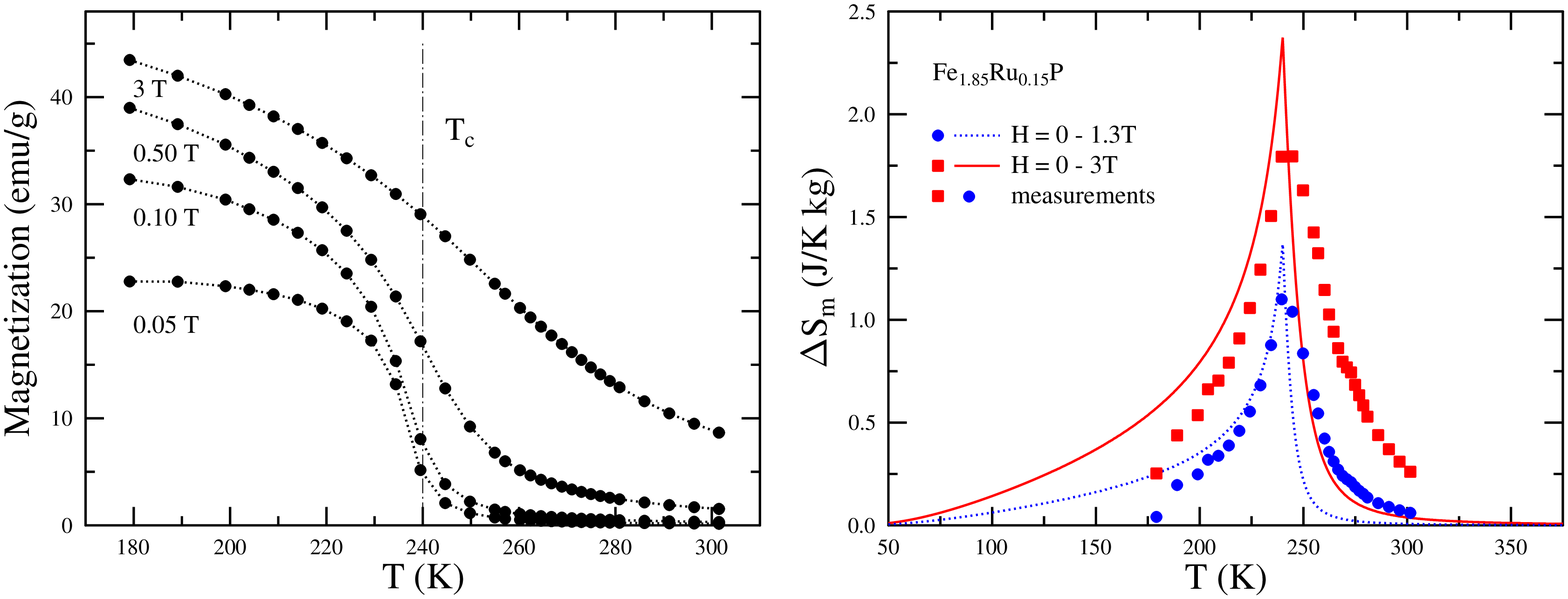}\caption{\label{fig-ferup} Magnetisation curves and entropy jump for Fe$_{1.85}$Ru$_{0.15}$P. In right panel circles and squares represent experimental data, whereas solid and dotted lines correspond to the calculated curves.}
\end{figure}

\begin{figure}[htb]
\includegraphics[width=1.0\textwidth]{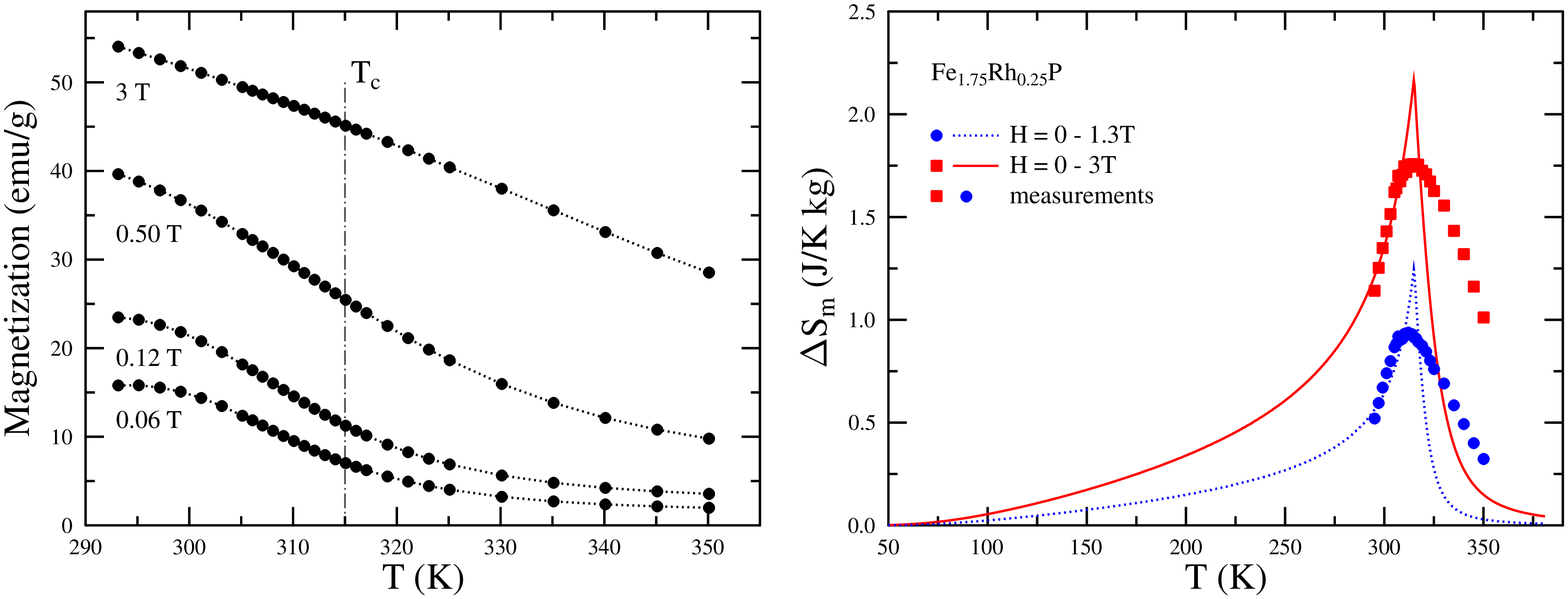}\caption{\label{fig-ferhp} Magnetisation curves and entropy jump for Fe$_{1.75}$Rh$_{0.25}$P. In right panel circles and squares represent experimental data, whereas solid and dotted lines correspond to the calculated curves.}
\end{figure}

The magnetic entropy changes, obtained from magnetisation measurements using the Maxwell-Weiss relation, are shown in \fref{fig-ferup} and \fref{fig-ferhp} for fields up to 3 T. Moreover, the comparison of Fe$_2$P and substituted compounds is presented in \fref{fig-entr}. We can observe, that with respect to pure Fe$_2$P, the magnitude of the measured entropy change is about twice smaller ($\Delta H =$ 1.3 T), i.e. 1.1 J/(kg.K) (Ru substitution) and 0.94 J/(kg.K) (Rh substitution). The calculated magnetic entropy changes, using the mean-field model and KKR-CPA ferromagnetic moments, correctly reproduce the variations determined experimentally. The computations tend to slightly overestimate the entropy jump (for $H =$ 0 - 1.3 T, the calculated $\Delta S_m$ are: 1.3 J/(kg.K) for the Ru case and 1.2 J/(kg.K) for the Rh case). This is in contrast to the Fe$_2$P compound, where the simulated $\Delta S_m$ is smaller than the measured one. This is probably due to the fact, that Fe$_2$P clearly exhibits first order magnetic transition, which gives sharper change of magnetisation (but less temperature extension), and consequently, when entropy is calculated via the Maxwell relation, the larger derivative $\partial M/\partial T$ gives the larger entropy jump. In both considered Fe$_{2-x}T_x$P alloys the change in $M(H, T)$ near $T_C$ is not so rapid (suggesting second order transition), thus the entropy curves are smoother than those observed for Fe$_2$P (see, Ref.~\cite{fruchart05}).

\section{Conclusions}

We have presented results of electronic structure calculations for Fe$_{2-x}T_x$P systems, obtained by the Korringa-Kohn-Rostoker method with the coherent potential approximation (KKR-CPA). To simulate the paramagnetic state of Fe$_2$P, the concept of disordered local moments (DLM) has been successfully applied. We have found that the magnetic moment appearing on Fe(3f) in ferromagnetic state completely vanishes in the DLM state, while that one on Fe(3g) remains only slightly changed.

The evolution of electronic structure, magnetic and magnetocaloric properties upon substituting Fe by Ru and Rh has also been studied, both theoretically (KKR-CPA calculations) and experimentally (magnetisation measurements). The pyramidal site preference upon substitution was found from the total energy analysis, in fair agreement with “selection rules”. Both Ru and Rh substituted in Fe$_2$P decrease the total magnetisation of the compounds, with stronger effect in the case of $T$ = Ru.
The electronic entropy was calculated as a function of temperature, using the computed KKR-CPA DOS’s. The change in magnetic entropy induced by the magnetic field was considered within the mean-field approximation. In spite of the fact, that the mean-field model was applied to a complex chemically disordered ferromagnetic system (two magnetically different sublattices), the agreement between theoretical model and experimental results was found to be quite satisfactory. 

\ack
This work was partly supported by Polish Ministry of Science and Higher Education under the grant no. P03 B 113-29 and 44/N-COST/2007/0 as well as the Polish-French collaboration project POLONIUM.

\section*{References}

\providecommand{\newblock}{}

\end{document}